\documentclass[aps,prb,twocolumn,showpacs,superscriptaddress,floatfix]{revtex4-1}
\usepackage{amssymb,amsmath}
\usepackage{graphicx}
\usepackage{float}
\usepackage{bm}
\usepackage{multirow}
\usepackage{dcolumn}
\usepackage{tabularx}
\usepackage{longtable}
\usepackage[dvipsnames,usenames]{color}

\newlength{\dbarheight}

\begin{document}
\title{Electric field control of Jahn-Teller distortions in bulk perovskites}

\author{Julien Varignon} \thanks{These two authors contributed
  equally} \affiliation{Physique Th\'eorique des Mat\'eriaux,
  Universit\'e de Li\`ege (B5), B-4000 Li\`ege, Belgium}
\author{Nicholas C. Bristowe} \thanks{These two authors contributed
  equally} \affiliation{Physique Th\'eorique des Mat\'eriaux,
  Universit\'e de Li\`ege (B5), B-4000 Li\`ege, Belgium}
\affiliation{Department of Materials, Imperial College London, London
  SW7 2AZ, UK} \author{Philippe Ghosez} \affiliation{Physique
  Th\'eorique des Mat\'eriaux, Universit\'e de Li\`ege (B5), B-4000
  Li\`ege, Belgium}

\date{\today}


\begin{abstract}  
The Jahn-Teller distortion, by its very nature, is often at the heart
of the various electronic properties displayed by perovskites and
related materials.  {\color{black}Despite the Jahn-Teller mode being
  non-polar in nature, we devise and demonstrate in the present letter
  an electric field control of Jahn-Teller distortions in bulk
  perovskites.}  The electric field control is enabled through an anharmonic 
lattice mode coupling between the Jahn-Teller distortion and a polar
mode.  We confirm this coupling, and explicitly an electric field
effect, through first principles calculations.  The coupling will
always exist within the $Pb2_1m$ space group, which is found to be the
favoured ground state for various perovskites under sufficient tensile
epitaxial strain.  Intriguingly, the calculations reveal that this
mechanism is not only restricted to Jahn-Teller active systems,
promising a general route to tune or induce novel electronic
functionality in perovskites as a whole.

\end{abstract}
\pacs{}
\maketitle

Perovskites, and related materials, are fascinating
systems exhibiting a diverse collection of properties, including
ferroelectricity, magnetism, orbital-ordering, metal-insulator phase
transitions, superconductivity and
thermoelectricity~\cite{zubko2011interface}.  
Despite the wide range of physical behaviour, a common point at the origin of many of them can be
identified as being the Jahn-Teller
distortion~\cite{Jahn-Teller-effect,Goodenough-JT}.
The Jahn-Teller distortion is itself intimately
linked to electronic degrees of freedom, since traditionally
it manifests to remove an electronic degeneracy, opening a band gap and favouring a 
particular orbital ordering, which in turn can affect magnetic ordering.  
Furthermore it plays an important role, for example, in
colossal magnetoresistance phenomena in doped manganites
~\cite{Raveau2001CMR},
superconductivity~\cite{RevModPhys.60.585-JT2,han2003strongTc-JT} or
the strong electronic correlation observed in the thermoelectric
NaCoO$_2$ family~\cite{berthelot2011electrochemicalNaCoO2}.

It would be highly desirable, for device functionality for example, to
be able to tune the Jahn-Teller distortion and hence its corresponding
electronic properties, with the application of an external electric
field. {\color{black}However, Jahn-Teller distortions are non-polar and
  hence not directly tunable with an electric field.}
One strategy to achieve an electric field control of Jahn-Teller
distortions might begin by engineering a polar structure.  Usually,
perovskites adopt a non-polar $Pbnm$ ground state, resulting in a
combination of three rotations of the oxygen octahedra ($a^-a^-c^+$ in
Glazer's notation~\cite{Glazer}), also called antiferrodistortive
(AFD) motions. Unfortunately, these AFD distortions are known to often
prevent~\cite{Benedek-fewFE} the appearance of the polarization in the
material.  The use of anharmonic lattice mode couplings between polar
and non-polar lattice distortions is a promising pathway to engineer
``improper" ferroelectricity in layered perovskite
derivatives~\cite{PTO-STO,NaLaMnWO6,HIF-Rondinelli,revue-novelMF}, and
furthermore to achieve novel functional possibilities via these
couplings~\cite{BFO-LFO,MOF2-Picozzi,MOF-Picozzi-PSSRRL2014,Ca3Mn2O7-Fennie,Benedek-polar-rotations,Vanadates-SL,Titanates-Nick}.
Interesting anharmonic couplings do not only exist in layered
perovskites, but can also appear in bulk ABO$_3$
perovskites~\cite{SRO-Naihua,Vanadates-SL,Nickelates-Rondinelli-amplimodes,Howard-JT1,BFO-OrbitalOrdering,PhysRevLett.112.057202,zhou2013strain}

Following this spirit, achieving an electric field control of
Jahn-Teller distortions necessarily requires the identification of a
material exhibiting
a relevant coupling between the polarization and the Jahn-Teller
distortion, which to the best of our knowledge has not yet been
discovered in bulk perovskites~\cite{ReviewDalton}. In the present letter we identify
such conditions, and demonstrate explicitly an electric field control,
in bulk perovskites using a combination of symmetry analysis and first
principles calculations.

\section*{Results}
\begin{figure}
\begin{center}
\resizebox{6.5cm}{!}{\includegraphics{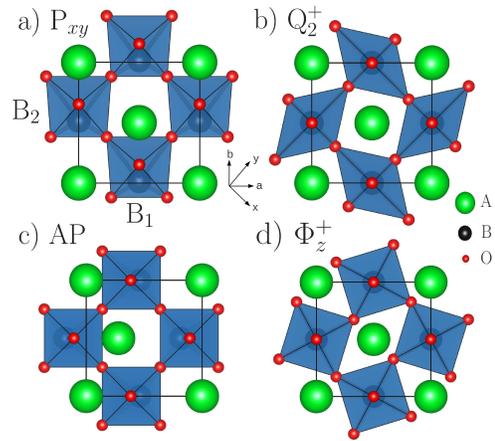}}
\end{center}
\caption{Schematic view of the four lattice distortions involved in
  the $Pb2_1m$ phase of  perovskites under tensile epitaxial strain. a)
  Polar distortion (irreps $\Gamma_5^-$) b) Q$_2^+$ Jahn-Teller
  distortion (irreps M$_3^+$) c) Anti-polar A distortion (irreps
  M$_5^+$) d) $a^0a^0c^+$ $\phi_z^+$ antiferrodistortive motion (irreps
  M$_2^+$).}
\label{f:dist}
\end{figure} 

The two required lattice distortions are pictured in
figure~\ref{f:dist}. a) and b). From a symmetry analysis of the different
tilt patterns in combination with possible Jahn-Teller lattice
distortions, Howard and Carpenter pointed out that a Jahn-Teller
distortion pattern automatically appears when considering the $Pbnm$
symmetry~\cite{Howard-JT1}, later explained in terms of anharmonic
couplings with AFD motions~\cite{Vanadates-SL}. As a consequence, a
Jahn-Teller distortion is not necessarily electronically driven, but can 
instead arise from lattice mode couplings in which case a
splitting of the electronic states may develop even in the absence of a
degeneracy of states. The present Jahn-Teller lattice motion corresponds
to a Q$_2$ mode as defined by Goodenough~\cite{Goodenough-JT},
corresponding to two B-O bond length contractions and two B-O
elongations. This motion orders at the M point of the Brillouin zone and
hence consecutive layers along the $\vec c$ axis of the $Pbnm$ phase
present in-phase distortions. Consequently, this motion is labelled
Q$_2^+$ (irreps M$_3^+$) throughout the whole manuscript.

Starting from the ideal $Pm\bar{3}m$ cubic perovskite phase, the condensation of
the polar mode P (irreps $\Gamma_5^-$) and the JT mode Q$_2^+$ lowers
the symmetry to a $Pb2_1m$ phase, a polar subgroup of $Pbnm$. We then
perform a free energy expansion~\cite{ISOTROPY} in terms of the
possible lattice distortions in this new phase and we identify, among
all the possible terms, some intriguing couplings:
\begin{eqnarray}
  \mathcal{F} &\propto &P Q_2^+ A +
  P^2 Q_2^+ \phi_z^+  + P \phi_z^+ A + Q_2^+ \phi_z^+A^2 
\label{e:free}
\end{eqnarray}
In this phase, the first two terms of equation~\ref{e:free} provide a
link between the polarization and the Jahn-Teller distortion.
These terms also involve two additional distortions: one anti-polar A motion
pictured in figure~\ref{f:dist}. c) and one $a^0a^0c^+ $ AFD motion
(labelled $\phi_z^+$) pictured in figure~\ref{f:dist}. d). Among all the
terms, the lowest order trilinear term of the form $P Q_2^+ A$
provides the desired direct coupling between the polarization and the JT
distortion. Thus, acting on the polarization with an external electric
field may modify the amplitude of the JT motion, and therefore all
related electronic properties.

However, as previously discussed, the $Pb2_1m$ symmetry is not the
common ground state in bulk perovskites~\cite{noteTrilinear}. Strain
engineering, through thin film epitaxy for example, can provide a
powerful tool to unlock a polar mode in
perovskites~\cite{revue-novelMF,SrTiO3-FERT,Strain-review,CaMnO3-Ph,CaMnO3-Prellier,EuTiO3-Fennie,EutiO3-Exp}. This
is the case for BiFeO$_3$ that was recently proposed to adopt an
unusual $Pb2_1m$ symmetry under large epitaxial tensile
strain~\cite{BFO-OrbitalOrdering,PhysRevLett.112.057202,fan2014structural}. This
particular phase was shown to develop both polar, anti-polar and
$a^0a^0c^+$ AFD motions~\cite{BFO-OrbitalOrdering}, which were later
demonstrated to be coupled together through the third term of
eq.~\ref{e:free}~\cite{PhysRevLett.112.057202}. Amazingly, the authors
reported the existence of an orbital ordering of the Fe$^{3+}$ $3d$
orbitals, explained from the coexistence of the polar and the
anti-polar motion yielding a particular lattice distortion
pattern~\cite{BFO-OrbitalOrdering}. This orbital-ordering is unusual
since in this system no Jahn-Teller effect is required to form a Mott
insulating state (Fe$^{3+}$ are in a half filled - high spin
$t_{2g}^3$$e_g^2$ configuration).  A Jahn-Teller effect or distortion
are yet to be reported in the $Pb2_1m$ phase of BiFeO$_3$ to the best
of our knowledge. From our symmetry analysis, we clearly demonstrate
that as this $Pb2_1m$ develops the three aforementioned distortions
(P, A and $\phi_z^+$), the free energy of eq.~\ref{e:free} is
automatically lowered through the appearance of a fourth lattice
distortion: a Jahn-Teller Q$_2^+$ motion.  Therefore, whilst it may
not itself be unstable, the Jahn-Teller motion is forced into the
system via this ``improper" mechanism arising from the trilinear
coupling~\cite{PTO-STO}.  This result clarifies the origin of the unusual
orbital-ordering displayed by BiFeO$_3$ and moreover, it provides a
pathway to achieve an electric field control of the orbital-ordering
in bulk perovskites.

The predicted highly strained $Pb2_1m$ phase in bulk perovskites is
not restricted to BiFeO$_3$, and it was predicted to occur also in
some titanates (CaTiO$_3$ and EuTiO$_3$)~\cite{BFO-OrbitalOrdering},
in BaMnO$_3$~\cite{BFO-OrbitalOrdering} and even in a Jahn-Teller
active compound TbMnO$_3$~\cite{TbMnO3-Pmc21}. The highly strained
bulk perovskites are then an ideal playground to demonstrate our
coupling between the polarization and the Jahn-Teller distortion. In
order to check the generality of our concept, we propose in this
letter to investigate several types of highly strained perovskites on
the basis of first principles calculations: i) non magnetic (NM)
SrTiO$_3$ ($t_{2g}^0 e_g^0$); ii) magnetic BaMnO$_3$~\cite{noteBMO}
($t_{2g}^3 e_g^0$) and BiFeO$_3$ ($t_{2g}^3 e_g^2$); iii) Jahn-Teller
active YMnO$_3$ ($t_{2g}^3 e_g^1$).

\begin{table}[b]
\centering
\begin{tabular}{ccccccc}
  \hline
  \hline
  & & SrTiO$_3$ & BaMnO$_3$ & BiFeO$_3$ & YMnO$_3$ \\
  strain & (\%) & +7.35~\cite{noteStrain-ElectricJahn-Teller} & +6.1~\cite{noteStrain-ElectricJahn-Teller} & +5.8~\cite{noteStrain-ElectricJahn-Teller} & +4.0~\cite{noteStrain-ElectricJahn-Teller}\\
  magnetism && NM & FM & AFMG & AFMG \\
  \multirow{2}{*}{P ($\Gamma_5^-$)}    & (\AA) & 0.615 & 0.421 & 0.346 & 0.753 \\
  & ($\mu C.cm^{-2}$) & 76   & 45 & 29 & 7~\cite{notePolarisationYMO}  \\
  Q$_{2}^+$ (M$_3^+$) & (\AA) & 0.232 & 0.190 & 0.644 & 0.737 \\
  A (M$_5^+$) &(\AA) & 0.558 & 0.217 & 1.072 & 0.940 \\
  $\phi_z^+$ (M$_2^+$)  &(\AA) & 0.640 & 0.059 & 1.668 & 1.733 \\
  gap &(eV) & 3.02 & 0.28 & 1.88 & 1.88 \\
  \hline
  \hline
\end{tabular}
\caption{Epitaxial strain (\%), magnetic ground state, amplitudes
  of distortions (\AA) and electronic band gap value (eV) 
 for each material. We
  emphasize that only the relevant distortions are summarized in the
  present table.}
\label{t:dist}
\end{table}

\begin{figure}
\centering
\resizebox{8.6cm}{!}{\includegraphics{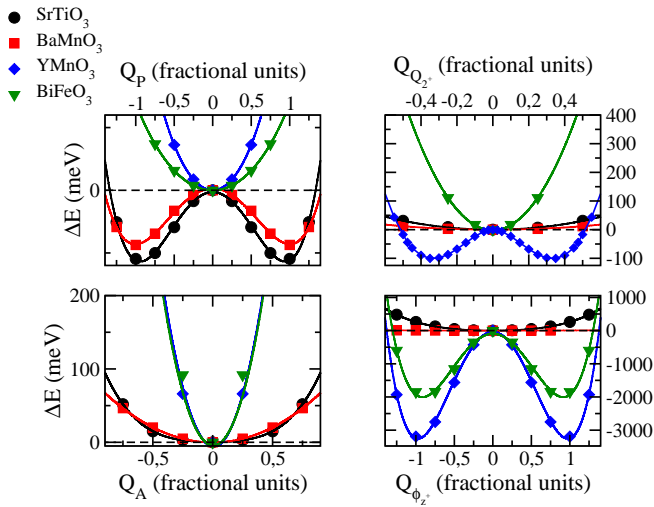}}
\caption{Potentials with respect to the amplitude of distortions of the
  four lattice motions producing the required $Pb2_1m$ for SrTiO$_3$ (black
  filled circles), BaMnO$_3$ (red filled squares), YMnO$_3$ (blue filled
  diamonds) and BiFeO$_3$ (green filled triangles) starting from the ideal
  $P4/mmm$ phase.} 
\label{f:potentials}
\end{figure}


We begin by investigating the possibility of a $Pb2_1m$ ground state
under large epitaxial tensile strain (the growth direction is along the [001]
axis of the $Pbnm$ structure).
Beyond around 5\% tensile strain, the four compounds indeed develop
the desired $Pb2_1m$ ground state.  Strained BaMnO$_3$ (ferromagnetic
FM) and YMnO$_3$ (G-type antiferromagnetic AFMG) exhibit a different
magnetic ground state compared to the bulk (AFMG and E-type
antiferromagnetic - $\uparrow \uparrow \downarrow \downarrow$ zig-zag
chains coupled antiferromagnetically along the $\vec c$ axis -
respectively) while BiFeO$_3$ (G-type antiferromagnetic AFMG) remains
in its bulk magnetic ground state.  We then perform a symmetry mode
analysis with respect to a hypothetical $P4/mmm$ phase (corresponding
to $Pm\bar{3}m$ for unstrained bulk compounds) in order to extract the
amplitude of the relevant lattice
distortions~\cite{noteAmplimodes-ElectricJahn-Teller} (see
table~\ref{t:dist}).  As expected, the four materials develop the
required distortions, and amazingly, the magnitude of the Q$_2^+$
Jahn-Teller distortion is relatively large, being for instance of the
same order of magnitude as the one developed in the prototypical
Jahn-Teller system LaMnO$_3$ (around 0.265
\AA~\cite{LaMnO3-Bousquet}). Additionally, the values of the
spontaneous polarization are rather large, reaching 76 $\mu C.cm^{-2}$
for SrTiO$_3$ for instance. Despite being highly strained, all
materials remain insulating, adopting reasonable electronic band gap
values (see table~\ref{t:dist}).

\begin{figure}
\centering
\resizebox{7.0cm}{!}{\includegraphics{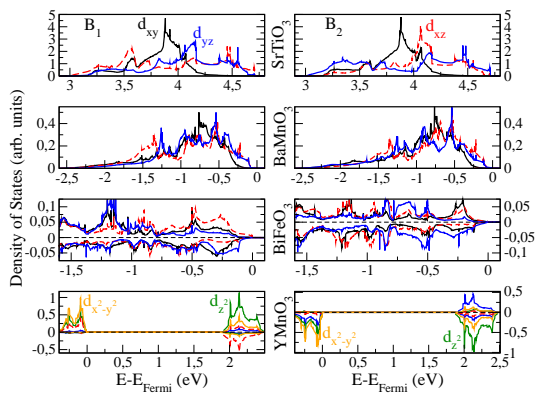}}
\caption{ Projected density of states on the $d$ levels on two
  neighboring B sites in the $(ab)$-plane of SrTiO$_3$, BaMnO$_3$,
  BiFeO$_3$ and YMnO$_3$. Local axes of the orbitals are displayed on
  figure~\ref{f:dist}. The Fermi level is located at 0~eV.}
\label{f:doss-tous}
\end{figure}

To shed more lights on the origin of this unusual $Pb2_1m$ phase we
compute the energy potentials with respect to the four distortions by
condensing individually each modes in an hypothetical $P4/mmm$ phase
(see figure~\ref{f:potentials}). Surprisingly, the appearance of the
$Pb2_1m$ phase is rather different for the four materials.  SrTiO$_3$
and BaMnO$_3$ only exhibit a polar instability, producing an $Amm2$
symmetry, consistent with previous reports of a polar phase for these
two materials under tensile
strain~\cite{SrTiO3-FERT,BaMnO3-Spaldin-PRB79-205119-2009}. Computing
the phonons in this particular $Amm2$ symmetry, only one hybrid
unstable phonon mode is identified for these two materials, having a
mixed character between the A, $\phi_z^+$ and Q$_2^+$ distortions,
 For BiFeO$_3$ and YMnO$_3$, the $a^0a^0c^+$ AFD motion is already
 unstable, which is expected since the $Pb2_1m$ symmetry for these two
 systems is derived from their bulk $R3c$/$Pbnm$
 phases~\cite{noteAmplimodes-ElectricJahn-Teller}. Additionally, the
 JT lattice distortion is also unstable in the $P4/mmm$ phase of
 YMnO$_3$ and appears as an electronic
 instability~\cite{noteElectronicInstabilityYMO}, which is expected
 since YMnO$_3$ is known to be Jahn-Teller active in the bulk. We
 emphasize at this stage that the polar mode in BiFeO$_3$ (and
 YMnO$_3$) is not unstable and {\color{black}therefore highly strained
   BiFeO$_3$ appears as an improper ferroelectric in contradiction to
   reference~\onlinecite{PhysRevLett.112.057202}}. Computing
 the phonons in the intermediate strained $Pbnm$ phase of both
 BiFeO$_3$ and YMnO$_3$ compounds reveals only one hybrid unstable
 mode, having a mixed character between P and A distortions.
Despite the apparent universal stability of this highly strained polar
phase, the mechanism yielding it is suprisingly different between the
compounds.

Regarding the electronic structure, we checked for the appearance of
an orbital ordering as observed in
BiFeO$_3$~\cite{BFO-OrbitalOrdering}. For the four compounds we report
the projected density of states on the $d$ levels of two neighboring B
sites in the $(ab)$-plane (see figure~\ref{f:doss-tous}).  For
SrTiO$_3$, a splitting of the $t_{2g}$ states, and especially between
the $d_{xz}$ and $d_{yz}$ orbitals, located at the bottom of the
conduction band arises. For BaMnO$_3$ and BiFeO$_3$, a similar
splitting between the $t_{2g}$ levels is observed near the Fermi
level, even if it is less pronounced for BaMnO$_3$ since it has the
smallest $Q_2^+$ distortion. Finally, YMnO$_3$ displays an orbital
ordering of the $e_g$ levels with predominantly $d_{x^2-y^2}$
occupation. This splitting is known to result of the Jahn-Teller
distortion in this A$^{3+}$Mn$^{3+}$O$_3$ class of
material~\cite{noteoo-YMO}. Additionally, an orbital ordering of the
$t_{2g}$ levels is occurring both in the conduction and the valence
bands.  To prove that the Jahn-Teller distortion, and not another
motion, is solely responsible for the orbital ordering we have
condensed all the modes individually and studied the density of states
(see supplementary figure 1).


\begin{figure}
\centering
\resizebox{6.5cm}{!}{\includegraphics{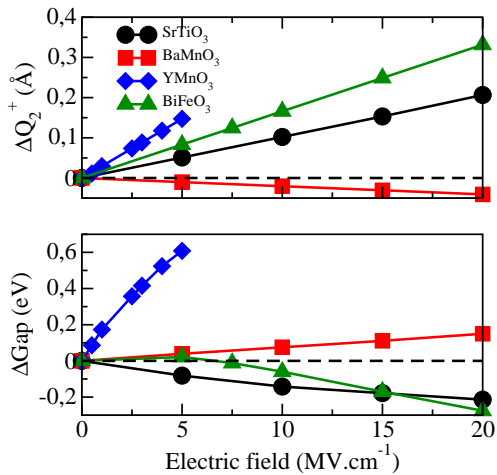}}
\caption{Relative electric field effect on the amplitude of the
  Jahn-Teller distortion (top panel) and the electronic gap value
  (bottom panel) on the four different compound.
}
\label{f:Efield}
\end{figure}

Finally, we explicitly demonstrate the electric field control of
Jahn-Teller distortion by computing, as a function of the electric
field $\vec E$ applied along the direction of the spontaneous
polarization, the evolution of both the JT distortion and the band
gap. Results are displayed in figure~\ref{f:Efield}.  The Jahn-Teller
distortion is effectively altered by the application of an electric
field along the polar axis through the first and second terms of
equation~\ref{e:free}. As the electric field increases, the amplitude
of the JT distortion is either amplified or decreased, being
renormalized to around 175\% for SrTiO$_3$ for an electric field
around 20 MV.cm$^{-1}$. The largest effect is however reached for
YMnO$_3$ which displays a renormalization of 130\% under moderate
electric field (around 5 MV.cm$^{-1}$). Therefore, this
renormalization of the JT distortion has consequences for instance on
the electronic band gap value, with an opening/closure around 0.6 eV
for YMnO$_3$ or 0.25 eV for SrTiO$_3$. It is then possible, through
the coupling between the polarization and the Jahn-Teller distortion
to act on the electronic band gap, and one may imagine a possible
control of a Metal-Insulator phase transition, optical properties, or
photovoltaic efficiency, for instance.


In conclusion, we have demonstrated in the  highly
strained $Pb2_1m$ phase of bulk perovskites the existence of a
coupling between a polar mode and the Jahn-Teller
distortion. This improper anharmonic coupling, established on
universal symmetry arguments, enables an electric field control of the
Jahn-Teller distortion, even in the case of non electronically
Jahn-Teller active systems. The generic mechanism may open novel
functionalities in perovskites as it will have consequences on related
electronic properties as proposed in the present letter. For instance,
such couplings may allow the tuning of Metal-Insulator phase
transitions, the control of orbital orderings, optical properties, or of the photovoltaic
efficiency. 

\section*{Methods}
First-principles calculations were performed with the VASP
package~\cite{VASP1,VASP2}. We used the PBEsol~\cite{PBEsol}+U
framework as implemented by Lichtenstein's method~\cite{LDAU-Lich}
(see the supplementary material for a discussion on the choice of the
U and J parameters). The plane wave cut-off was set to 500 eV and we
used a 6$\times$6$\times$4 k-point mesh for the 20 atom $Pb2_1m$
phase. PAW pseudopotentials~\cite{PAW} were used in the calculations
with the following valence electron configuration: $3s^23p^64s^2$
(Sr), $4s^24p^65s^2$ (Ba), $4s^24p^65s^24d^1$ (Y), $6s^26p^3$ (Bi),
$3p^64s^23d^2$ (Ti), $3p^64s^23d^5$ (Mn), $3p^64s^23d^6$ (Fe) and
$2s^22p^4$ (O). Spontaneous polarizations were computed using the
Berry-phase approach and phonons and Born effective charges were
computed using the density functional perturbation
theory~\cite{RMP-Baroni}. The electric field effect was modelled using
a linear response approach by freezing-in some lattice distortion into
the system~\cite{MethodeJorge,JVYMnO3}. Symmetry mode analyses were
performed using the Amplimodes software from the Bilbao
Crystallographic server~\cite{Amplimodes1,Amplimodes2}.

\acknowledgments Work supported by the ARC project TheMoTherm and
F.R.S-FNRS PDR project HiT4FiT. Ph. Ghosez acknowledges the Francqui
Foundation and N.C. Bristowe the Royal Commission of the Exhibition of
1851 for a fellowship at Imperial College London. Calculations have
been performed within the PRACE projects TheoMoMuLaM and
TheDeNoMo. They also took advantage of the C\'eci facilities funded by
F.R.S-FNRS (Grant No 2.5020.1) and Tier-1 supercomputer of the
F\'ed\'eration Wallonie-Bruxelles funded by the Waloon Region (Grant
No 1117545).



\begin{thebibliography}{53}%
\bibitem{zubko2011interface}
 P. Zubko, S. Gariglio, M. Gabay, P. Ghosez, and J.-M.
Triscone, Annu. Rev. Condens. Matter Phys. {\bf 2}, 141 (2011).

\bibitem{Jahn-Teller-effect}
H. K\"oppel, D. R. Yarkony, and H. Barentzen, {\em The Jahn-
Teller Effect} (Springer, 2009).

\bibitem{Goodenough-JT}
 J. Goodenough, Annual review of materials science {\bf 28}, 1 (1998).

\bibitem{Raveau2001CMR}
 B. Raveau, M. Hervieu, A. Maignan, and C. Martin, J.
Mater. Chem. {\bf 11}, 29 (2001).

\bibitem{RevModPhys.60.585-JT2}
 J. G. Bednorz and K. A. M\"uller, Rev. Mod. Phys. {\bf 60}, 585 (1988).

\bibitem{han2003strongTc-JT}
 J. Han, O. Gunnarsson, and V. Crespi, Phys. Rev. Lett. {\bf 90}, 167006 (2003).

\bibitem{berthelot2011electrochemicalNaCoO2}
 R. Berthelot, D. Carlier, and C. Delmas, Nature Materials {\bf 10}, 74 (2011).

\bibitem{Glazer} A. Glazer, Acta Cryst.  B {\bf 28}, 3384 (1972).

\bibitem{Benedek-fewFE}
N.A . Benedek and C. J. Fennie, J. Phys. Chem. C {\bf 117}, 13339 (2013).


\bibitem{PTO-STO}
E. Bousquet, M. Dawber, N. Stucki, C. Lechtensteiger,
P. Hermet, S. Gariglio, J. M. Gariglio, and P. Ghosez,
Nature {\bf 452}, 732 (2008).

\bibitem{NaLaMnWO6}
 T. Fukushima, A. Stroppa, S. Picozzi, and J. M. Perez-
Mato, Phys. Chem. Chem. Phys. {\bf 13}, 12186 (2011).

\bibitem{HIF-Rondinelli}
J. M. Rondinelli and C. J. Fennie, Adv. Materials {\bf 24}, 1961
(2012).

\bibitem{revue-novelMF}
 J. Varignon, N. C. Bristowe, E. Bousquet, and P. Ghosez,
Comptes Rendus Physique {\bf 16}, 153 (2015).

\bibitem{BFO-LFO}
 Z. Zanolli, J. C. Wojdel, J. I\~{n}iguez, and P. Ghosez, Phys.
Rev. B {\bf 88}, 060102(R) (2013).

\bibitem{MOF2-Picozzi}
 A. Stroppa, P. Barone, P. Jain, J. M. Perez-Mato, and
S. Picozzi, Adv. Mater. {\bf 25}, 2284 (2013).

\bibitem{MOF-Picozzi-PSSRRL2014}
Y. Tian, A. Stroppa, Y.-S. Chai, P. Barone, M. Perez-
Mato, S. Picozzi, and Y. Sun, Physica Status Solidi (RRL) (2014).

\bibitem{Ca3Mn2O7-Fennie}
N. A. Benedek and C. J. Fennie, Phys. Rev. Lett. {\bf 106},
107204 (2011).

\bibitem{Benedek-polar-rotations}
N. A. Benedek, A. T. Mulder, and C. J. Fennie, J. Solid
State Chem. {\bf 195}, 11 (2012).

\bibitem{Vanadates-SL}
J. Varignon, N. C. Bristowe, E. Bousquet, and P. Ghosez,
(2014), arXiv preprint arXiv:1409.8422.

\bibitem{Titanates-Nick}
 N. C. Bristowe, J. Varignon, D. Fontaine, E. Bousquet,
and P. Ghosez, Nat. Commun. {\bf 6}, 6677 (2015).

\bibitem{SRO-Naihua}
 N. Miao, N. C. Bristowe, B. Xu, M. J. Verstraete, and
P. Ghosez, J. Phys.: Cond. Matt. {\bf 26},
035401 (2014).

\bibitem{Nickelates-Rondinelli-amplimodes}
 P. V. Balachandran and J. M. Rondinelli, Phys. Rev. B {\bf 88}, 054101 (2013).


\bibitem{ReviewDalton}
N. A. Benedek, J. M. Rondinelli, H. Djani, Ph. Ghosez and PH. Lightfoot, Dalton Transactions (2015).

\bibitem{noteTrilinear} Being in the $Pb2_1m$ phase is sufficient but
  {\em a priori} not mandatory to achieve our goal. From
  eq~\ref{e:free}, only the coupling between the polar and JT
  distortions is key to achieve an electric field control of the
  Jahn-Teller distortions. Indeed, one can imagine a metastable phase
  only developing the antipolar A distortion. Applying an electric
  field would activate the polar mode and through the first term of
  eq.~\ref{e:free}, the Jahn-Teller distortion may automatically
  appear. However, in practice this metastable phase does not seem to
  be favoured, and being in the ground state ferroelectric phase may
  bring added functionality such as switchable behaviour.

\bibitem{Howard-JT1}
M. A. Carpenter and C. J. Howard, Acta Cryst. B {\bf 65}, 134 (2009).

\bibitem{BFO-OrbitalOrdering}
 Y. Yang, W. Ren, M. Stengel, X. Yan, and L. Bellaiche,
Physical Rev. Lett. {\bf 109}, 057602 (2012).

\bibitem{PhysRevLett.112.057202}
 Y. Yang, J.  \'I\~{n}iguez, A.-J. Mao, and L. Bellaiche, Phys.
Rev. Lett. {\bf 112}, 057202 (2014).

\bibitem{zhou2013strain}
 Q. Zhou and K. M. Rabe, arXiv preprint arXiv:1306.1839
(2013).

\bibitem{ISOTROPY}
 D. M. Hatch and H. T. Stokes, J. Applied Cryst. {\bf 36}, 951 (2003).

\bibitem{SrTiO3-FERT}
 J. H. Haeni, P. Irvin, W. Chang, R. Uecker, P. Reiche, Y. L.
Li, S. Choudhury, W. Tian, M. E. Hawley, B. Craigo, A. K.
Tagantsev, X. Q. Pan, S. K. Streiffer, L. Q. Chen, S. W.
Kirchoefer, J. Levy, and D. G. Schlom, Nature {\bf 430}, 758
(2004).

\bibitem{Strain-review}
 J. Junquera and P. Ghosez, J. Comput. Theo. Nanosci. {\bf 5},
2071 (2008).

\bibitem{CaMnO3-Ph}
 S. Bhattacharjee, E. Bousquet, and P. Ghosez, Phys. Rev.
Lett. {\bf 102}, 117602 (2009).

\bibitem{CaMnO3-Prellier}
 T. G\"{u}nter, E. Bousquet, A. David, P. Boullay, P. Ghosez,
W. Prellier, and M. Fiebig, Phys. Rev. B {\bf 85}, 214120
(2012).

\bibitem{EuTiO3-Fennie}
 C. J. Fennie and K. M. Rabe, Phys. Rev. Lett. {\bf 97}, 267602
(2006).

\bibitem{EutiO3-Exp}
 J. H. Lee, L. Fang, E. Vlahos, X. Ke, Y. W. Jung, L. F.
Kourkoutis, J.-W. Kim, P. J. Ryan, T. Heeg, M. Roeck-
erath, V. Goian, M. Bernhagen, R. Uecker, P. C. Hammel,
K. M. Rabe, S. Kamba, J. Schubert, J. W. Freeland, D. A.
M. C. J. Fennieand, P. Schiffer, V. Gopalan, E. Johnston-
Halperin, and D. G. Schlom, Nature {\bf 466}, 954 (2010).

\bibitem{fan2014structural}
Z. Fan, J. Wang, M. B. Sullivan, A. Huan, D. J. Singh,
and K. P. Ong, Scientific Reports {\bf 4} (2014).

\bibitem{TbMnO3-Pmc21}
Y. Hou, J. Yang, X. Gong, and H. Xiang, Physical Rev. B {\bf 88}, 060406 (2013).

\bibitem{noteBMO} While the ground state of BaMnO$_3$ has been shown to
  adopt a hexagonal polar $P63cm$ structure~\cite{BMO-improper}, BaMnO3 can also be
  stabilized with a perovkite form under tensile strain~\cite{BaMnO3-Spaldin-PRB79-205119-2009} .


\bibitem{noteStrain-ElectricJahn-Teller} The strain percentage is
  defined as the elongation with respect to the
  $\sqrt{2}a$,$\sqrt{2}b$ axes of the pseudo cubic structure
  corresponding to the fully relaxed ground state. The fully relaxed
  ground states for all compounds are given in the supplementary
  materials.

\bibitem{noteAmplimodes-ElectricJahn-Teller} We only report in table I
  the relevant distortions for the proposed mechanism. One should
  notice that due to a small tolerance factor, BiFeO$_3$ and
  YMnO$_3$xs still develop large $a^−a^-c^0$rotations in their ground
  state, in addition to other antipolar modes. Their $Pb2_1m$ phase
  may appear to be derived from a $R3c$ or $Pbnm$ structure
  respectively.

\bibitem{LaMnO3-Bousquet} J. H. Lee, K. T. Delaney, E. Bousquet,
  N. A. Spaldin, and K. M. Rabe, Physical Rev. B {\bf 88}, 174426
  (2013).

\bibitem{BaMnO3-Spaldin-PRB79-205119-2009}
 J. M. Rondinelli, A. S. Eidelson, and N. A. Spaldin, Phys. Rev. B {\bf 79}, 205119 (2009).



\bibitem{noteElectronicInstabilityYMO} The JT distortion in YMnO$_3$
  appears through an electronic instability mechanism in the high
  symmetry phase.  Indeed, only removing the symmetry on the
  electronic wavefunction while keeping the centrosymmetric positions
  for the cations produces already an energy gain, that is then
  amplified by the resulting JT lattice distortions.

\bibitem{noteJorge} In
  reference~\onlinecite{PhysRevLett.112.057202}, authors report a
  relatively weak polar instability in the $P4/mmm$ phase that we do
  not obtain in our simulations. This contradiction may be related to
  technical details or the magnitude of the strain applied, but in any
  case will not affect the final result of the present letter.

\bibitem{noteoo-YMO} The two average ab-plane Mn-O bond lengths are
  evaluated to be $\left<d_{Mn−O}\right>_{ab,1}$= 1.903 \AA~and
  $\left<d_{Mn−O}\right>_{ab,2}$= 2.438 \AA~ while the average Mn-O
  bond length along the c axis is around $\left<d_{Mn−O}\right>_{c}$=
  1.900 \AA. Consequently, the $d_{x^2−y^2}$ orbital should be more
  stable than the $d_{z^2}$ orbital. Considering the sole Q$_2^+$
  distortion, the two in-plane Mn-O bond lengths are 2.294 \AA~ and
  1.772 \AA~ while the out of plane bond length is 1.769 \AA.


\bibitem{BMO-improper} J. Varignon and P. Ghosez, Phys. Rev. B {\bf
  87}, 140403(R) (2013).

\bibitem{VASP1}
 G. Kresse and J. Haffner, Phys. Rev. B {\bf 47}, 558 (1993).

\bibitem{VASP2}
 G. Kresse and J. Furthm\"{u}ller, Computational Materials
Science {\bf 6}, 15 (1996).

\bibitem{PBEsol}
J. P. Perdew, A. Ruzsinszky, G. I. Csonka, O. A. Vydrov,
G. E. Scuseria, L. A. Constantin, X. Zhou, and K. Burke,
Phys. Rev. Lett. {\bf 100}, 136406 (2008).

\bibitem{LDAU-Lich}
 A. I. Liechtenstein, V. I. Anisimov, and J. Zaanen, Phys.
Rev. B {\bf 52}, R5467 (1995).

\bibitem{PAW}
P. E. Bl\"{o}chl, Physical Review B {\bf 50}, 17953 (1994).

\bibitem{RMP-Baroni}
 S. Baroni, S. de Gironcoli, A. Dal Corso, and P. Giannozzi,
Rev. Mod. Phys. {\bf 73}, 515 (2001).

\bibitem{MethodeJorge}
J. \'I\~{n}iguez, Phys. Rev. Lett. {\bf 101}, 117201 (2008).

\bibitem{JVYMnO3}
J. Varignon, S. Petit, A. Gell\'e, and M. B. Lepetit, J.
Phys.: Condens. Matt. {\bf 25}, 496004 (2013).

\bibitem{Amplimodes1}
 D. Orobengoa, C. Capillas, M. I. Aroyo, and J. M. Perez-
Mato, J. App. Cryst. {\bf 42}, 820 (2009)

\bibitem{Amplimodes2}
 J. Perez-Mato, D. Orobengoa, and M. Aroyo, Acta Cryst. A {\bf 66}, 558 (2010).


\end{thebibliography}
\end{document}